# Smart Electromechanical Pumping of Electrons in a Nanopillars Transistor

Yue-Min Wan and Heng-Tien Lin

*Department of Electronic Engineering, I-Shou University, Kaohsiung Taiwan 84001, Republic of China*

**Abstract**

Analysis of room-temperature current-voltage (*I-V*) characteristics of a silicon box in a nanopillar transistor suggests that a weak electromechanical coupling λ ~ 0.17 is responsible for the stable tunnel of single-electron. The dynamics involves a few electrons and the numbers (N) specified are periodical at 3, 6, and 12. Quantized currents are observed at N = 7 and 13, indicating that the box is a man-made atom. At a large λ ≥ 0.5, instability however dominates the *I-V* by showing interference, channel closure and the change of tunnel direction. Overall, the interplay of even and odd electrons between different channels also shows that the box operates itself like a smart quantum pump.

Recently, the coupling of single-electron tunnel [1] to mechanical vibration in various single-electron transistors (SETs) such as single oscillating molecules [2], suspended semiconductor beams [3], and carbon nanotubes [4] has attracted much attention targeting at the issue of how mechanical feedback can influence electrical transport. It is well known that the transport is featured by a series of isolated peaks in the *I-V* characteristics [5-6] and the spacing between peaks defines the Ec, while prior to the charging Coulomb blockade (CB) [7] strictly prohibits current to flow. Recent studies, however, have projected new features that could appear near the CB; Gorelik *et al*. [8] first pointed out that shuttle effect can occur, where the central box can oscillate and its amplitude will reach a stable limit at a large voltage bias. Soon after, Blanter *et al*. [9] demonstrated that this effect can further create noises with some evidences being reported in Refs. [10-12]. Clearly, these studies have highlighted the importance of understanding new phenomena in SETs disregarding the possibility of adding some complexity to data analysis.

Aside from these works, a critical question in our view remains unanswered. That is to what degree the feedback can modify electron tunnel. Here, we address this issue through detailed measurements of *I-V* characteristics of the newly developed nanopillar SET where its size can be tuned with precision [13]. We find that at a weak electron-phonon interaction, the first excited state of $N = 3$ correspond to one free electron in each degree of freedom, and the states of $N = 6$ and 12 are stable with even number in each degree. Introduction of the seventh electron at $N = 7$ initiates a dynamical pumping of electrons ($\Delta N = 1$) through the double-barrier box and the entrance of the thirteen electron at $N = 13$ creates a staircase by doubling the quantized current. Together, these numbers leads us to the first finding of a long-sought man-made atom. At a strong interaction, the distribution of electrons becomes uneven, leading to a strong competition between different channels. The self-organized interplay of even and odd electrons, however, always finds a way to stabilize the box thus making it operating also like a smart quantum pump.

The SET as shown in Fig. 1(a) was fabricated on a p-type wafer and the central Si box was isolated from the top and bottom electrodes by a nitride barrier. Fabrication details of the device can be found in Ref. [13]. Post-examination using ellipsometer shows that the overall structure is ~ SiNx-3.5nm/Si-3nm/SiNx-3.5nm. The nitride is known of having a potential barrier 2.1 eV which is sufficient to excite only a few electrons. An Al (300 nm) side gate was arranged on the side (9 nm apart) to control the box potential. Transport measurements were conducted in a probe station by using a three-terminal HP 4156 C, which has 1mV and 10 fA resolutions in an ambient environment of 300K.

As presented in Fig. 2, periodical peaks clearly dominate the central *I-Vs* and suggest a stable tunnel of electron in the SET. The charging energy $E_c = e^2/2C$ derived from the experimental settings of $C = \varepsilon_r\varepsilon_o A/D$ ~ 2.2 aF, $A = L \times H = 8\times 8$ nm$^2$, $\varepsilon_o = 8.85\times 10^{-12}$ C$^2$/Nm$^2$, $\varepsilon_r = 11.7$, and $D = 3$ nm of ~ 35 mV agrees with the peak splitting very well in confirming the excellence of the device.

Once the behavior of single-electron tunnel is confirmed, then we decide the electronic-state by employing the well-known single-particle formula in Eq. (1). As denoted (arrow) in Fig. 2(a), the threshold $V_t$ ~ 0.24 V corresponds to the doublet states [3,2,2] and [2,3,2] (see Table I), meaning that the capital N (nx+ny+nz) equals seven and the distribution of the small n is quite uniform. This finding is very important as the same kind of results have also been discovered in several previous occasions [13-14].

$$E(n_x, n_y, n_z) = \frac{\pi^2 \hbar^2}{2m}(\frac{n_x^2}{L^2} + \frac{n_y^2}{W^2} + \frac{n_z^2}{H^2})  \qquad (1)$$

The unique combination of n = 3 in the x-channel (or y) plus n =2 in z is believed to contain the key ingredients for a stable tunnel [15]. Based on this fact, we therefore propose a dynamical model. Because of the softness in all materials at such a high temperature, they ought to be elastic and will subject to deformation under the presence of an external force-field [16]. In response, the box will vibrate and become an oscillator. Specifically, in Fig. 3, at the initiation of $V_{ds}$, the nitride on the right (or the left) will bend due to an electron-phonon interaction. The free electron of nz =1, Fig. 3(a), will be excited to a higher state $E_2$, which in turn will mediate another force on the left barrier (right) to yield a displacement ΔX as illustrated in Fig. 3(b). The ΔX will then create a downward current pulse $\Delta I_{ds}$ as illustrated in the inset. As expected, the resilience in SiNx can force itself to bounce back and squeezes the box in further, Fig. 3(c), thus creating a sharp flip to the opposite end as $\Delta I_{ds}'$ in Fig. 3(d). After the completion of these cycles, a zig-zag signal is then created [17]. Blending with some uncertainty in the energy of electron due to a much shorter life time (< $10^{-12}$ sec) [14], the signal can become asymmetric, i.e, $\Delta I_{ds} \neq \Delta I_{ds}'$ as evidenced by the giant noises below $V_t$. In a lower temperature [18], these noises are expected to become much shaper and then turn into a complete Coulomb blockade as the rigidity will make them disappear in supporting the zero-power-average $e\int I_{ds} dV_{ds} = 0$ as detected.

Quantitatively, one can estimated these mechanically-induced noises by assuming $\Delta I_{ds}$ = $e\Delta V n_e \Gamma \approx 3$ pA, where Γ is the tunneling rate [19], ΔV is the total volume affected, $n_e$ is the induced charge density ∝ AΔX/3, and A is the effective tunnel area. Given A = 64 nm$^2$, e = 1.6 x$10^{-19}$ C, Γ = $10^{12}$/sec and N = $10^{16}$/cm$^3$, ΔX is found to be ~ 3 Å. The work is K(ΔX)$^2$/2, where K is the spring constant of SiNx and it equals σA (σ is the elastic stress in unit of Pa/m$^2$) [20].

Given a mean value ~ 200 G, one finds K ~ 0.16 N/m and the work done is ~ 50 meV. Note that this value agrees with the energy of the first excited state [1,1,1] very well, thus pointing to two interesting implications; first, it means that the condition for setting up a 3D box vibration is via the excitation of single-electron in each degree of freedom; second, this work also severs as the upper bound of the total energy Ee that can be stored in the resonator. The Ee will be released in the transition from [1,1,1] to [2,2,2] as the resonator stop to vibrate and it creates a significant amount of negative/counter current flow [21] to offset the $I_{ds}$ after an extra ordinarily large noise at [1,2,2]/[2,1,2] ~ 196 meV .

At the state of nz =2, the wall vibrations in Figs. 4 will be limited to the lateral directions only; as the top layer bends down, it can discharge an electron out of the box via Coulomb repulsion. As the oxide reaches its balance, in Fig. 4(c), the electron on the left will move forward to the right and leave a hole behind [22]. This hole, moment later, can draw an electron from the charge reservoir as the oxide reaches its maximum to the top. Such kind of partial charging is also favored by the lowering of the Fermi level $E_F$ to allow an easy entrance of the succeeding electron [23]. Such scheme also explains why a stable tunnel requires the exact mixture of electrons in the box in order to synchronize a constant energy exchange in between electronics and mechanics.

Notice that just before [3,2,2] threshold, the configuration is [2,2,2] which exactly satisfies the classical theorem of equal partition. At this point, the entrance of the next, i.e, the "seventh" electron triggers a two-level resonance. The adding of this *magic* entity apparently will go into the easiest channel and that will make the gap ΔEn ~ 30 meV (see Table I) separating the higher N = 7 state and the lower N = 6 state as the genetic charging energy. However, to fulfill a mechanical feedback, the Ee has to be taken into account. By matching ΔEn + ΔEe to the Ec, a ΔEe ~ 5 meV is therefore determined. The small ratio ~ 1/10 of ΔEe/Ee thus confirms that the feedback required is indeed small in yielding a much smaller ΔX ~ 1 Å [24] as predicted by the shuttling effect, where a fast damping in the oscillating amplitude as the state is raised from 1 to 3. The $\lambda = \Delta Ee/\Delta En \sim 0.17$ also confirms that the coupling is weak. Interestingly enough, beyond this point the power transferred becomes positive and $I_{ds}$ will be limited by a RC constant, which is estimated to be ~ $10^{-7}$s for a total junction resistance ~ $10^{11}$ Ω and a capacitance ~ $10^{-18}$ aF. With these numbers, the peak $I_{ds}$ = e/RC calculated is ~ $10^{-12}$ A in good accordance with the data.

The two-level resonance amazingly continues without disruption until it comes to the [5,4,4] at ~ 910 meV. This state again satisfies the same kind of even/odd mixture of N = 13 and the state as the second magic number making a discrete jump in $I_{ds}$ by doubling its magnitude but without changing its periodicity. The step increase as an essential character of a staircase signifies that electron-excitation remains in the same box, however, the package delivered becomes 2*e* (see the central inset of Fig. 4) at the cost of introducing 1*e*. Note that the jump is irreversible; in Fig. 2(b), one clearly sees that as $V_{ds}$ decreases, the drop from 2e to e is at [4,4,4] ~ 857 meV, which obviously is a clever choice in terms of mechanical balance. At $V_{ds}$ further below the disappearance of noises suggests that the box retains its best mode and this unusual state-sensitive behavior seems to explain some of the most intriguing phenomena observed in hysteretic *I-V*. Most important, the reminiscence of the atomic p-shell like order in ΔN = 6 (from 13 to 7) suggests that the SET manifests itself like a man-made atom.

By contrast, in the reversed charging as shown in Fig. 5(a), the resonance somehow becomes unstable and it is due to the bypass of [2,2,2]. Note that the threshold recognized is at the doublets [3,2,1]/[2,3,1] and obviously they are inferior to the [2,2,2] for N = 6. As a result, the adding of the seventh electron, which is supposed to be into nz to stabilize the box, goes into nx instead, making the resonance unstable. Furthermore, the instability continues by yielding the following sequence [3,2,1], [4,2,1], [4,3,1], [5,2,1], [4,4,1], [6,1,1], [6,2,1], [6,3,1] for the first eight peaks. The unbalanced profile in between nx and ny is very evident and at some point the deviation of $\Delta n = |nx-ny|$ is as large as 5, making interference inevitable as evidenced by the strong modulation in $I_{ds}$ [25-26]. Since all the nz are the same of 1, the contribution from mechanical feedback is expected to be high. As estimated, the $\Delta Ee$ will increase at least by a factor of three, leading to a large $\lambda \geq 0.51$.

Intuitively, this 2D instability should be suppressed by an increasing bias, but to our surprise, another complication springs; notice that as the charging comes to the state [4,4,1] ~ 230 meV, another series of small satellite peaks appears, meaning that the transition from nz = 1 to 2 occurs. The three-way competitions eventually make the $I_{ds}$ drop quickly and once it comes across [4,4,2] ~ 350 meV, the channel is closed by showing zero current. The meet of this all-even state appears to be a destiny because the box has to stabilize itself first before reopening and this is indeed observed at [5,4,2] ~ 408 meV, where another series of broader peaks resume. Although it is not clear to us which factor makes such a drastic change, it is speculated that it has to do with the overall structures of the SET or with a slight deviation in the thickness of SiNx.

Finally, the nature of 3D shows its effect on the side charging. In Fig. 5(b), one sees that electron begins to charge into the box at $V_{gs}$ ~ 1 V with a periodicity $\Delta V_{gs}$ of ~ 0.5V. Given the identities of $e = \alpha \Delta V_{gs} C_g$ ($\alpha$ gate-dot coupling strength) and $C_g = \alpha C$, we finds $\alpha$ ~ 0.37 and $C_g$ ~ 0.83 aF. The $\alpha$ fits into the empirical plot in Ref. [13] very well for an oxide of ~ 9 nm and confirms above analysis. The total energy acted on the box in this specific measurement is $e(V_{ds} + \alpha V_{gs}) = 420$ meV, which is found corresponding to the singlet [2,2,3]. Notice that this state is right above the [5,4,2] indicating that the closure of the channel does provide an opportunity for electron to change its moving direction [27]. Overall, the interplay of even and odd electrons between different channels also makes the SET operating like a smart quantum pump.

*Acknowledgements* – We thank NDL for SET fabrication, Chih-An Chen and Prof. Hsiang-Chen Hsu (ISU) for numerical analysis using finite-element method. The work was supported by NSC (Republic of China) under the Contract No. NSC95-2112-M-214-001.

Figure Captions

Fig. 1 (a) SEM picture of the nanopillar transistor. (b) Schematic drawing of the 3D vertical device.

Fig. 2 $I_{ds}$ versus $V_{ds}$ at $V_{gs}$ =0. (a) is the charging from 0 to -1 V and the asteroid marks the state of [3,2,1]. (b) shows the process of discharging.

Fig. 3 Box vibration and zig-zag noises from the state of n =1. (a) to (d) illustrates the cycled vibrations due to single-electron force field. Central inset is the generated zig-zag noise. In case the box is squeezed from both sides as (c), the electron can excite to $E_2$. Similar vibrations are also expected for n = 3.

Fig. 4 Pumping of electron under the assist of side-wall vibrations. (a)-(d) shows how individual electron can tunnel. The vibrations are originated from nx = 3. Inset shows ideal profiles of space charge for nz =1 to 4 [28]. The tunnel of nz = 4 is also recorded in the data.

Fig. 5 (a) $I_{ds}$ versus $V_{ds}$ for $V_{gs}$ = 0 V and (b) $I_{ds}$ versus $V_{gs}$ for $V_{ds}$ = 50 mV. Giant zig-zag noises are clearly registered in the bottom figure.

Table I Quantized states and their energies in the box of 8 x 8 x 3 nm$^3$. States on the left are identified in Figs. 2(a) and 2(b), while those on the right are identified in Figs. 5(a) and 5(b).

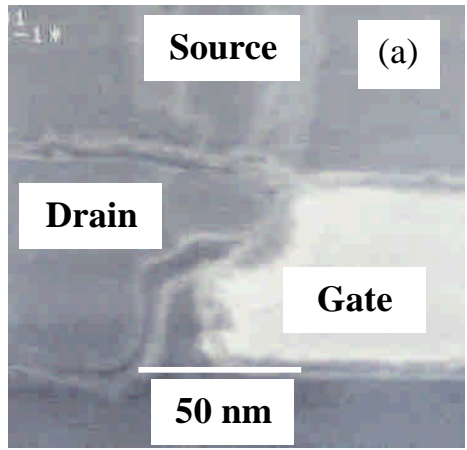

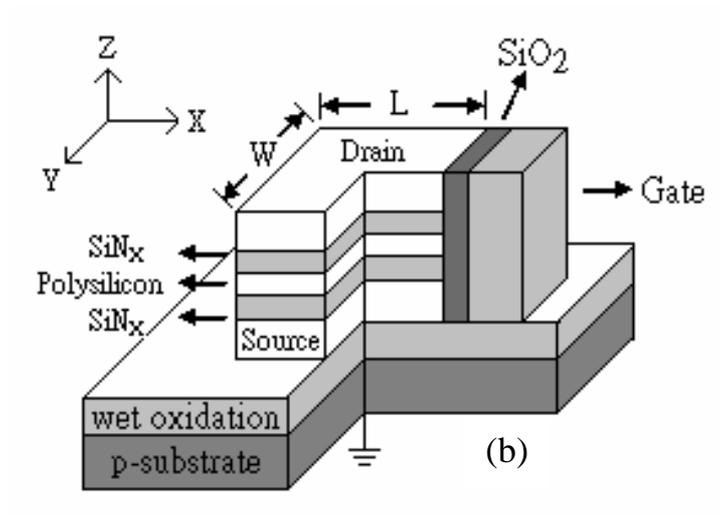

Fig. 1

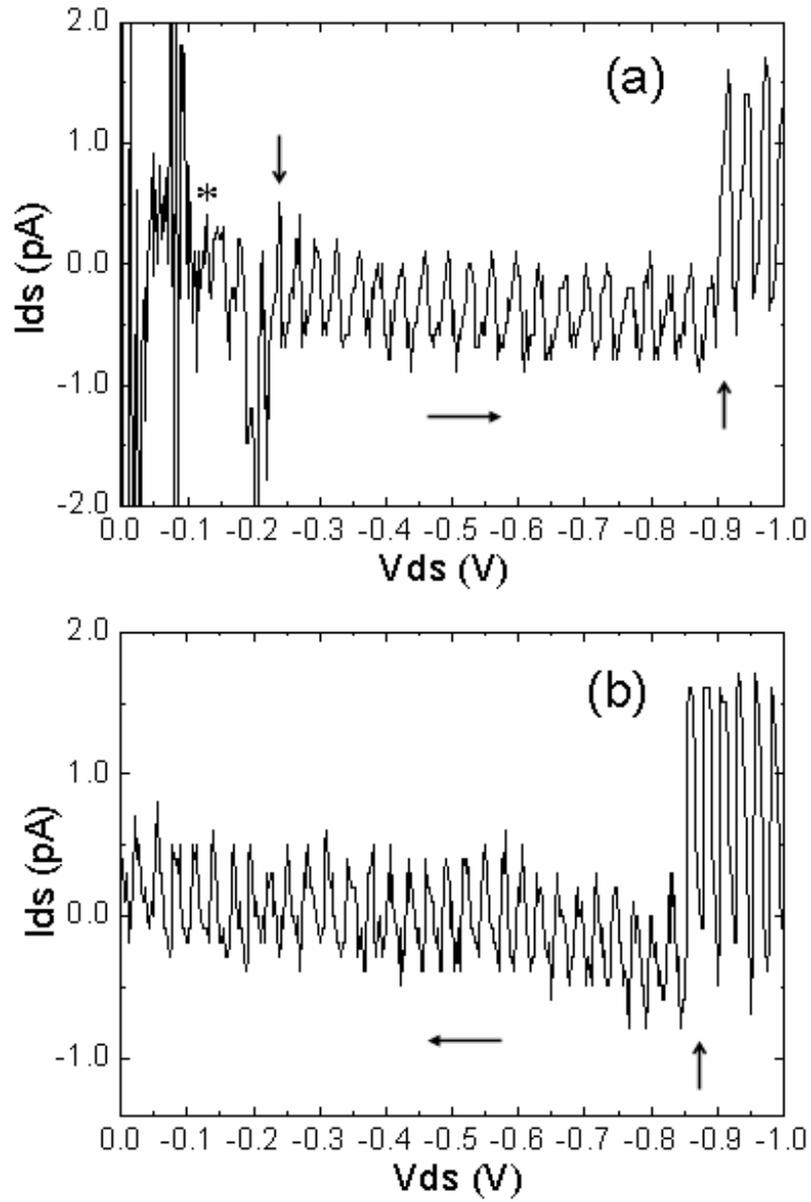

Fig. 2

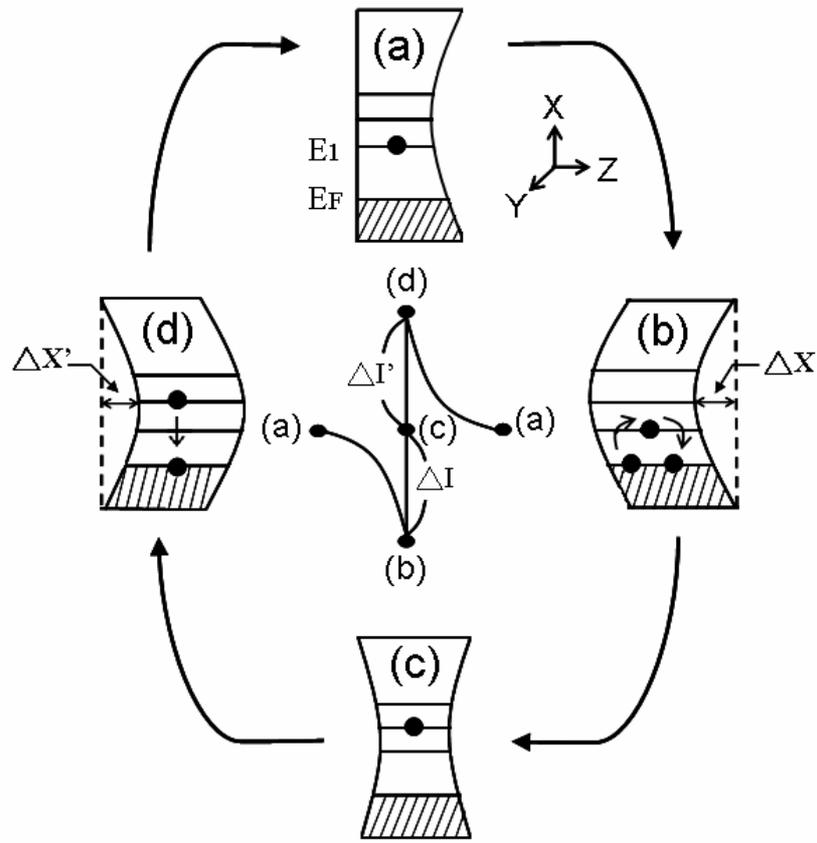

Fig. 3

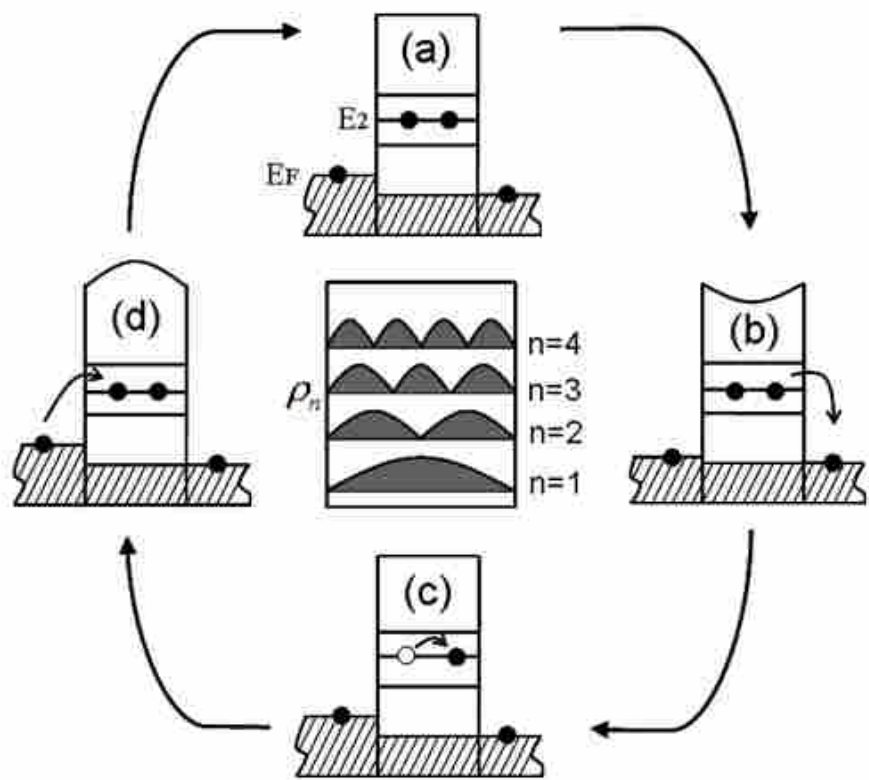

Fig. 4

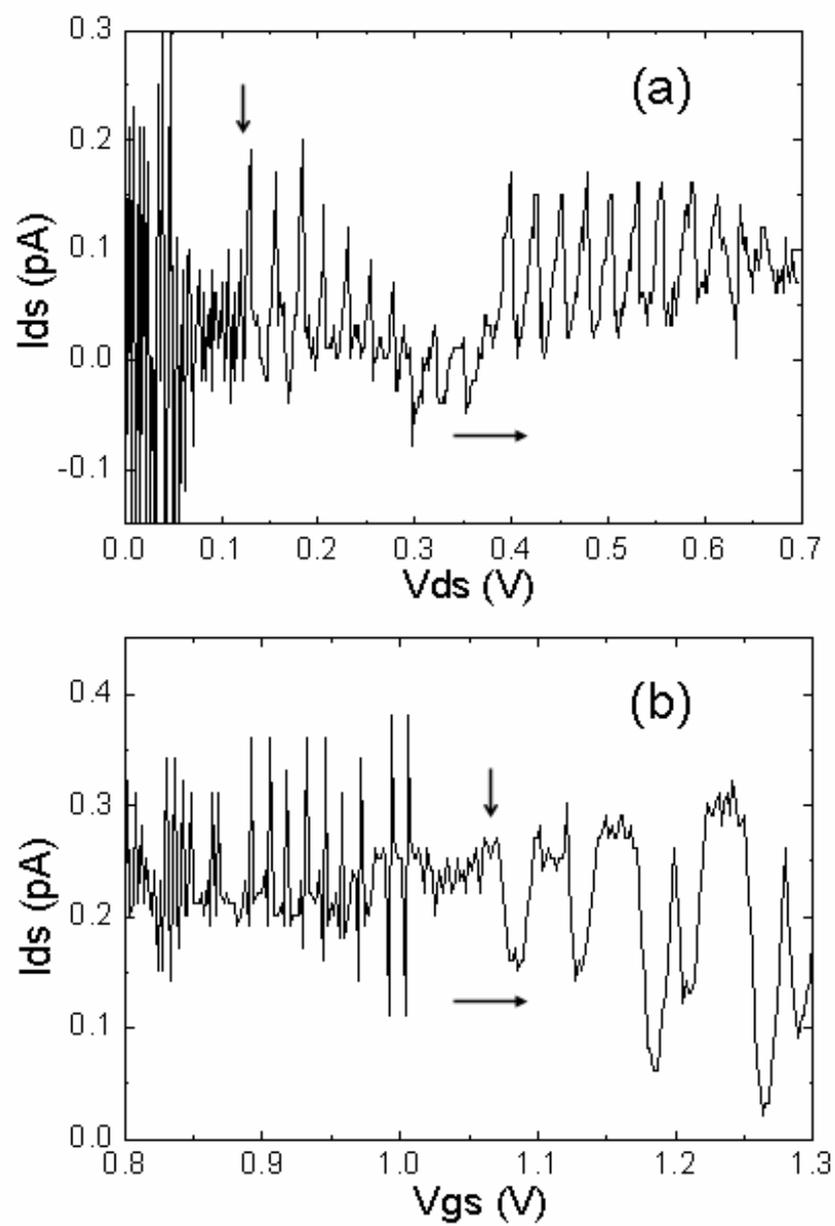

Fig. 5

Table I

| [ $n_x$, $n_y$, $n_z$ ] | $E_n$(meV) | [ $n_x$, $n_y$, $n_z$ ] | $E_n$(meV) |
|---|---|---|---|
| [ 1,1,1 ] | 53.5 | [ 2,4,1 ][ 4,2,1 ] | 159.3 |
| [ 2,3,1 ][ 3,2,1 ] | 118.2 | [ 3,4,1 ][ 4,3,1 ] | 188.7 |
| [ 2,2,2 ] | 214.2 | [ 2,5,1 ][ 5,2,1 ] | 212.2 |
| [ 2,3,2 ][ 3,2,2 ] | 243.6 | [ 4,4,2 ] | 355.2 |
| [ 4,4,4 ] | 856.7 | [ 5,4,2 ][ 4,5,2 ] | 408.1 |
| [ 4,5,4 ][ 5,4,4 ] | 909.6 | [ 2,2,3 ] | 423.1 |